\documentclass[aps,prl,showpacs,twocolumn]{revtex4}

\usepackage{amssymb}
\usepackage{epsfig}
\usepackage{graphicx}

\begin{document}

\title{Localized gap soliton trains of Bose-Einstein condensates in an optical lattice}

\author{D. L. Wang$^{1,2,3}$, X. H. Yan$^1$, W. M. Liu$^3$}
\address{$^1$College of Science, Nanjing University of Aeronautics
 and Astronautics, Nanjing, 210016, China}
\address{$^2$Department of Physics, Xiangtan University, Xiangtan, 411105, China}
\address{$^3$Beijing National Laboratory for Condensed Matter Physics, Institute of Physics,
Chinese Academy of Sciences, Beijing 100080, China}

\date{\today}

\begin{abstract}
We develop a systematic analytical approach to study the linear and
nonlinear solitary excitations of quasi-one-dimensional
Bose-Einstein condensates trapped in an optical lattice. For the
linear case, the Bloch wave in the $nth$ energy band is a linear
superposition of Mathieu's functions $ce_{n-1}$ and $se_n$; and the
Bloch wave in the $nth$ band gap is a linear superposition of $ce_n$
and $se_n$. For the nonlinear case, only solitons inside the band
gaps are likely to be generated and there are two types of solitons
-- fundamental solitons (which is a localized and stable state) and
sub-fundamental solitons (which is a lacalized but unstable state).
In addition, we find that the pinning position and the amplitude of
the fundamental soliton in the lattice can be controlled by
adjusting both the lattice depth and spacing. Our numerical results
on fundamental solitons are in quantitative agreement with those of
the experimental observation [Phys. Rev. Lett. {\bf92}, 230401
(2004)]. Furthermore, we predict that a localized gap soliton train
consisting of several fundamental solitons can be realized by
increasing the length of the condensate in currently experimental
conditions.
\end{abstract}


\maketitle

\section{I. Introduction}

Loading Bose-Einstein condensates (BECs) in an optical lattice
formed by a laser standing wave has received increasing interest in
the study of nonlinear atomic optics
\cite{Barrett,Orzel,Cataliotti,Greiner}. Understanding the
properties of BEC in an optical lattice is of fundamental importance
for developing novel application of quantum mechanics such as atom
lasers and atom interferometers \cite{5,6,7,8,9,9a,9b}.
Theoretically, some approximation methods are borrowed from solid
state physics, which are used to investigate the dynamics of this
system. It is mainly due to the fact that there are considerable
resemblances between BEC droplet localized in an optical lattice and
electron in a lattice. According to the theory of solid state
physics, there exist band gaps between adjacent energy bands in the
band structure of solid. In general, the energy bands exhibits
spatially oscillating phenomena. As discussed in Refs. \cite{10} and
\cite{11}, however, it is possible to generate soliton in the band
gap when the nonlinearity compensates for atom dispersion caused by
inter-site tunneling. The band gap soliton can be called gap
soliton. The existence of the gap solitons was first predicted based
on coupled-mode theory \cite{12}, in analogy to optical gap solitons
in Bragg gratings. Such a prediction was validated by a number of
groups using some approximation approaches, such as tight binding
approximation \cite{10}, a complete set of on-site Wannier states
\cite{14}, an effective mass formula \cite{15} and plane wave method
\cite{16}. Although they provide a convenient way to study the gap
soliton of the BEC, the validity depends greatly on the nature of
the underlying problem. From this point of view, it is desirable to
develop a method that does not rely on above approximations
\cite{17}.

Strictly speaking, an accurate solution can be obtained by exactly
solving the full nonlinear Schr\"{o}dinger equation with a periodic
potential. However, it is very difficult to derive analytical
solutions because the full nonlinear Schr\"{o}dinger equation is
nonintegrable \cite{18}. Consequently, some asymptotic approaches
and numerical simulations are used to investigate this question.
Using multiple scale method, Konotop and Salerno \cite{17} predicted
that bright solitons could come into being in a BEC with a positive
scattering length and dark solitons could be stable with a negative
scattering length. Subsequently, these predictions were proved by
using asymptotic theories \cite{19,21}. Employing numerical
simulations, Louis et al. \cite{22} analyzed the existence and
stability of spatially extended (Bloch type) and localized states of
a condensate in the band gaps of the linear Bloch-wave spectrum.

Especially, Eiermann et al. \cite{23} reported that gap soliton do
neither move nor change their shape and atom numbers during
propagation. That is to say, the gap soliton is pinned in an optical
lattice without attenuation and change in shape. Such a soliton can
be regarded as a spatially localized gap soliton, which is also
called fundamental soliton in Ref. \cite{37}. More importantly,
localized bright solitons would be very useful for future
applications, such as atomic interferometry \cite{24}. Subsequently,
some explanations to this observation were proposed by using
numerical simulations (see Refs. \cite{25,26,27} and references
therein).

To better understand the characteristics of the linear and nonlinear
solitary excitations of quasi-one-dimensional (1D) BECs trapped in
an optical lattice, we develop a multiple scale method to derive
analytically an explicit expression of the wave function. It is
found that there are two types of gap solitons in the band gaps. One
is fundamental soliton, which is always stable and pins a fixed
position; the other is always unstable and decays gradually due to
losing a part of its atoms. The paper is organized as follows: In
Sec. II, we derive 1D amplitude and phase equations from the
original three-dimensional Gross-Pitaevskii (GP) equation.
Subsequently, by analyzing the stability regions of soliton
formation, we obtain the formation condition of the fundamental
solitons in the band gaps. A novel linear dispersion relation
arising from the ground state and sound speed of this system in the
band gaps are obtained in Sec. III. In Sec. IV, we develop a
multiple scale method to study the nonlinear dynamics of the system.
We derive a solution of the wave function and discuss its dynamical
stability in the band gaps. It is found that the pinning position
and amplitude of the fundamental solitons are controlled by
adjusting both lattice depth and spacing. Furthermore, we propose an
experimental protocol to observe a localized gap train consisting of
several fundamental solitons in the condensate under currently
experimental conditions. A brief summary is given in section V.

\section{II. EQUATIONS of AMPLITUDE AND PHASE}

Based on mean-field approximation, the time-dependent GP equation of
full BEC dynamics reads \cite{16,18,22}
\begin{equation}
i\hbar\frac{\partial\Psi}{\partial
T}=[-\frac{\hbar^2}{2m}\nabla^2+V(X,Y,Z)+g|\Psi|^2]\Psi,\label{GP1}
\end{equation}
where $\Psi(X,Y,Z,T)$ is the order parameter of condensate, and
$(X,Y)$ and $Z$ are the directions of strong transverse confinement
and lattice. $N=\int dr|\Psi|^2$ is the total number of atoms, and
$g=4\pi\hbar^2a_s/m$ is inter-atomic interaction strength with the
atomic mass $m$ and the $s$-wave scattering length $a_s$ ($a_s>0$
represents the repulsive interaction). The combined potential
$V(X,Y,Z)$ of the optical lattice and magnetic trap is
\begin{eqnarray}
V(R^2,Z)=E_0\sin^2(\frac{\pi Z}{d}) &+& \frac{1}{2}m(\omega^2_\bot
R^2 +\omega^2_Z Z^2),\nonumber\\
&& \omega_Z\ll\omega_\bot, \label{potential}
\end{eqnarray}
where $R^2=X^2+Y^2$, $E_0$ is the lattice depth. $d=\lambda_L/2$ is
the lattice spacing, where $\lambda_L$ is the wavelength of laser
beams. $\omega_Z$ and $\omega_\bot$ are frequencies of the magnetic
trap in the axial $(Z)$ and transverse ($X$ and $Y$) directions,
respectively. By introducing the dimensionless variables
$t=\omega_\bot T$, $(r,z)=a^{-1}_0 (R,Z)$ with transverse harmonic
oscillator length $a_0=\sqrt{\hbar/m \omega_\bot}$, and
$\psi=\sqrt{a^3_0/N}\Psi$, we obtain the following dimensionless GP
equation
\begin{eqnarray}
i\frac{\partial\psi}{\partial t}=-\frac{1}{2}\nabla^2 \psi+[V_0
\sin^2(\frac{\pi z}{D}) \nonumber\\ +\frac{1}{2}(r^2+\Omega^2
z^2)]\psi+Q|\psi|^2\psi, \label{GP2}
\end{eqnarray}
where $V_0=E_0/(\hbar \omega_\bot)$, $D=a^{-1}_0 d$,
$\Omega=\omega_z /\omega_\bot \ll 1$ and $Q=4\pi a_s /a_0$.
Expressing the order parameter in terms of modulus and phase, i.e.,
$\psi=\sqrt{n} \exp(i\Phi)$, and then separating real and imaginary
parts, we obtain
\begin{equation}
\frac{\partial n}{\partial t} +\nabla \cdot (n \nabla
\Phi)=0,\label{A1}
\end{equation}

\begin{eqnarray}
\frac{\partial \Phi}{\partial t} &+& V_0 \sin^2(\frac{\pi z}{D})
+\frac{1}{2}(r^2+\Omega^2 z^2)+\frac{1}{2} (\nabla \Phi)^2 \nonumber\\
&&-\frac{1}{2 \sqrt{n}}\nabla^2 \sqrt{n} +Q n=0.\label{A2}
\end{eqnarray}

Equations (\ref{A1}) and (\ref{A2}) are (3+1)-dimensional, nonlinear
and dispersive equations with a variable coefficient. To solve these
equations, we introduce some reasonable approximations. Considering
a $^{87}$Rb condensate in a cigar-shaped trap with the frequencies
of $\omega_z=2\pi\times0.5$ Hz and $\omega_\bot=2\pi\times85$ Hz
\cite{23}, we get $\Omega \approx 0.006$. The value of $\Omega$ is
so small that the variation of the profile of the order parameter is
slow in the $z$ direction. Thus, the wave function can be separated
by $\psi(r,z,t)=G_0 (r)\phi(z,t)$ with $\phi(z,t)=A(z,t)\exp[-i\mu
t+i\varphi(z,t)]$. Here, the modulus and the phase are
$\sqrt{n}=G_0(r) A(z,t)$ and $\Phi=-\mu t+\varphi(z,t)$
respectively. Owing to the strong confinement in the transverse
direction, the spatial structure of function $G_0(r)$ can be well
described by a solution of two-dimensional radial symmetric quantum
harmonic-oscillator equation, i.e., $\nabla^2_\bot G_0 +2G_0-r^2
G_0=0$. The ground-state solution has the form
$G_0(r)=C\exp(-r^2/2)$, where $C=1/\sqrt{\pi}$ can be found from the
normalization condition $\int^\infty_{-\infty}|G_0|^2 rdr=1$.
Substituting them into Eqs. (\ref{A1}) and (\ref{A2}), we obtain

\begin{equation}
\frac{\partial A}{\partial t}+\frac{\partial A}{\partial z}
\frac{\partial\varphi}{\partial z} +\frac{1}{2} A \frac{\partial^2
\varphi}{\partial z^2}=0,\label{A}
\end{equation}

\begin{eqnarray}
-\frac{1}{2} \frac{\partial^2 A}{\partial z^2} + [\frac{1}{2} (
\frac{\partial \varphi}{\partial z})^2-\mu +1 +\frac{\partial
\varphi}{\partial t} + \nonumber\\ V_0 \sin^2(\frac{\pi z}{D})]A
+Q' A^3=0,\label{P}
\end{eqnarray}
with $Q'=Q/(2\pi)=2a_s/a_0$. In order to obtain 1D amplitude and
phase equations, we have multiplied Eq. (\ref{P}) by $G^*_0$ and
then integrated the resulting equation once with respect to the
transverse coordinate to eliminate the dependence on transverse
plane. An approach similar to this one has been widely used in
quasi-1D (cigar-shaped) BEC problems \cite{22,28,29}.

\section{III. LINEAR BLOCH MODES}

We now consider a BEC trapped in a 1D optical lattice. Due to the
strong confinement in the transverse direction, the system is
similar to a wave guide, in which the excitation propagates in the
elongated direction \cite{31}. The strong confinement also ensures
the dynamical stability of the linear excitation \cite{31}.
Therefore, we set $A=u_0 (z) +\alpha (z,t)$ with $\alpha
(z,t)=\alpha_0 \exp(i\alpha_1) +c.c.$ and
$\varphi=\varphi_0\exp(i\alpha_1) +c.c.$. Here, $c.c.$ is complex
conjugate and $\alpha_1=kz-\omega t$. $k$ is the wave number and
$\omega$ is the eigenfrequency. Without loss of generality, we
assume $u_0 (z)$ characterizing the condensate background.
Considering that $\alpha_0$ and $\varphi_0$ are small constants, we
obtain
\begin{equation}
i\omega\alpha_0=i k\varphi_0\frac{\partial u_0}{\partial z}
-\frac{1}{2} k^2 u_0\varphi_0,\label{L1}
\end{equation}
\begin{equation}
(\mu-1) u_0=-\frac{1}{2}\frac{\partial^2 u_0}{\partial z^2} +V_0
\sin^2(\frac{\pi z}{D}) u_0 +Q'u^3_0,\label{L2}
\end{equation}
\begin{eqnarray}
(\mu-1)\alpha_0 &=& \frac{1}{2} k^2 \alpha_0 -i\omega\varphi_0 u_0
+V_0 \sin^2(\frac{\pi z}{D})\alpha_0 \nonumber\\
 &&+3Q'u^2_0 \alpha_0  \label{L3}
\end{eqnarray}
from the linearization of Eqs. (\ref{A}) and (\ref{P}). Under the
linear case, $Q'\approx0$. Equation (\ref{L2}) is turned into
Mathieu's equation \cite{32,33}
\begin{equation}
\frac{d^2 u_0}{d\eta^2}+[p+2q\cos(2\eta)]u_0=0,\label{M}
\end{equation}
with $\eta=\pi z/D$, $q=-V_0 D^2/(2\pi^2)$ and
$p=q+2D^2(\mu-1)/{\pi^2}$. Based on the Floquet-Bloch theorem, $u_0$
can be represented a superposition of Bloch waves, i.e., $u_0
(\eta)=b_1 \exp (i\nu\eta) u_{01}(\eta) +b_2 \exp (-i\nu\eta)
u_{02}(\eta)$, where $u_{01}(\eta)$ and $u_{02}(\eta)$ are Mathieu's
functions ($ce_n$ or $se_n$), $b_1$ and $b_2$ are arbitrary
constants, and $\nu$ is a Floquet exponent. If $\cos(\nu\pi)=\pm 1$,
the solutions of Mathieu's equation are periodic functions and can
be expanded as Fourier series (detailed expression in Ref.
\cite{33}).

It should be mentioned that in recent experiments the characteristic
lattice spacing $d$ is determined by the angle between the
intersecting laser beams forming the lattice and varies in the range
$0.4-1.6\mu m$ \cite{34}. The lattice depth $E_0$ scales linearly
with the light intensity, and varies between $0$ and
$E^{max}_0\approx20 E_{rec}$, where $E_{rec}=\pi^2\hbar^2/(2md^2)$
is the lattice recoil energy \cite{34}. So, the dimensionless
parameters $V_0$ and $D$ are in the range of $0< V_0\leq 7.0\times
10^2$ and $0.3\leq D\leq 4.05$. For convenience, we here set
$D=3.14$ in our calculation.

From the eigenvalues $p$ and $q$ of Mathieu's function (the
text-book analysis can be found, e.g., in Ref. \cite{32}), figure 1
presents the band-gap diagram for the extended solutions of Eq.
(\ref{M}) which describe noninteracting condensed atoms in an
optical lattice. The results are presented for the parameter domain
$(V_0, \mu)$ relevant to our problem. One can find that the energy
bands (shaded areas) are separated by the band gap regions. In these
band gaps, unbounded solutions exist. The band edges correspond to
exactly periodic solutions. The regions 1 and 2 represent the first
and second energy band respectively, while the regions $I$ and $II$
denote the first and second band gap respectively. From Fig. 1, we
can conclude that the Bloch wave $u_0$ in the $nth$ energy band is
the linear superposition of the Mathieu's functions $ce_{n-1}$ and
$se_n$, and $u_0$ in the $nth$ band gap is the linear superposition
of $ce_n$ and $se_n$. As is shown below, a complete band-gap
spectrum of the matter waves in an optical lattice provides
important clues on the existence and the stability regions of
solitons.

\begin{figure}[tbp]
\centering
\includegraphics[width=0.5\textwidth]{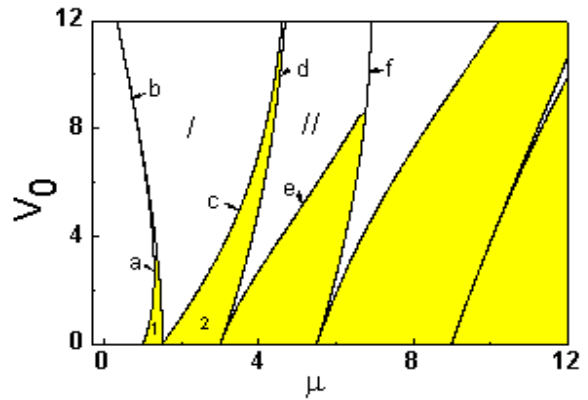}
\hspace{0.1cm} \caption{(Color online). Bloch band of BEC in an
optical lattice in the linear regime is functions of the optical
depth $V_0$ and the chemical potential $\mu$. The shaded areas
denote the energy bands. The regions 1 and 2 represent the first
and second energy band respectively. The areas $I$ and $II$
represent the lowest two band gaps in the spectrum. Band edges
(solid line) a, $b, c, d, e, f, \cdots$ correspond to the
eigenvalues and eigenstates $P^c_0 (ce_0), P^s_1 (se_1), P^c_1
(ce_1), P^s_2 (se_2), P^c_2 (ce_2), P^s_3 (se_3), \cdots$ of
Mathieu's equation, respectively.}
\end{figure}

We next consider the case of $Q'\neq 0$, and discuss the stability
problem of soliton formation. Utilizing Eqs. (\ref{L1})-(\ref{L3}),
we obtain

\begin{equation}
\omega^2=(\frac{1}{2u_0}\frac{\partial^2 u_0}{\partial z^2}
+\frac{1}{2}k^2+2Q' u^2_0)(\frac{ik}{u_0}\frac{\partial
u_0}{\partial z}+\frac{1}{2}k^2).\label{De}
\end{equation}
Setting $\omega=\omega_r +i\omega_i$ (where the subscripts denote
the real and imaginary parts) \cite{35}, one gets that $
\omega^2_r=(\gamma/4)\sqrt{k^4+(4 k^2/u^2_0) (\frac{\partial
u_0}{\partial z})^2} +(k^2 \gamma/4)$ and $
\omega^2_i=(\gamma/4)\sqrt{k^4+(4 k^2/u^2_0) (\frac{\partial
u_0}{\partial z})^2} -(k^2 \gamma/4)$, where $\gamma=[1/(2 u_0)]
\frac{\partial^2 u_0}{\partial z^2} +k^2/2 +2Q'u^2_0$. If the
imaginary part of quasi-particle frequency is a nonzero value, the
corresponding Bloch wave exhibits exponential growth and hence the
state $\psi$ is dynamical instability \cite{22}. If the frequency of
the associated quasi-particle spectrum is real, the soliton would be
stable. From the expression of the imaginary part of the frequency,
one can see the dependence of the instability growth rate $\omega_i$
on $k$ and $\frac{1}{u_0}\frac{\partial u_0}{\partial z}$. On the
one hand, when $k=0$, one finds $\omega=0$, which is an inessential
solution. On the other hand, it is impossible for
$\frac{1}{u_0}\frac{\partial u_0}{\partial z}$ being equal to zero
because $u_0$ is the Bloch wave in the energy bands or band gaps. If
$\frac{1}{u_0}\frac{\partial u_0}{\partial z}$ is a purely imaginary
number (also obtained from Eq. (\ref{De})), the wave function $\psi$
possesses dynamical stability. We therefore conclude that the stable
condition of soliton formation is $u_0=\exp(i\beta z)$ where $\beta$
is an arbitrary real constant. Because the Bloch wave in the $nth$
energy band is the linear superposition of the Mathieu's functions
$ce_{n-1}$ and $se_n$, it always not satisfies the stable condition.
Only if Bloch wave in the $nth$ band gap has the form of $u_0=\beta
ce_n +\beta i se_n$, soliton possesses dynamical stability. Thus the
linear dispersion relation of the $nth$ band gap is
\begin{equation}
\omega^2=(\frac{k^2}{2}-\frac{kn \pi}{D})(\frac{k^2}{2}-\frac{n^2
\pi^2}{2 D^2}+2Q'u^2_0).\label{De2}
\end{equation}

Under long-wave approximation, the sound speed is
\begin{equation}
V_g=\lim_{k\longrightarrow 0} \frac{\partial \omega}{\partial k} =
\pm \sqrt{Q' u^2_0-\frac{n^2}{4}(\frac{\pi}{D})^2},\label{V}
\end{equation}
where the positive (negative) sign represents the rightward
(leftward) propagation of the wave packets. For the case of
$D\longrightarrow\infty$, the external potential would be a harmonic
potential [see Eq. (\ref{potential})] and a corresponding sound
speed is $V_{gh}=\sqrt{Q'} u_0$ in our notation. This behavior is
consistent with both the experimental \cite{36} and theoretical
results \cite{29}. Obviously, the second term under the radical sign
in Eq. (\ref{V}) is arisen from the optical lattice potential. The
sound speed is the largest in the first band gap and gradually
decreases with $n$. Generally speaking, the value of $1/D$ in the
experiments \cite{23,34} would be larger than that of $Q'$. It
implies that the linear dispersion relation and sound speed are
dependent mainly on the lattice spacing.

\section{IV. NONLINEAR BLOCH MODES}

\subsection{A. The explicit expression of the wave function}

To better understand the nonlinear dynamics of BEC in an optical
lattice, we here develop a multiple scale method to derive an
explicit expression of the wave function of the condensates in an
optical lattice. By means of asymptotic expansion in nonlinear
perturbation theory, we propose that the amplitude and phase can be
expanded by multiple scale methods. In the case of that,
mathematically, any parameter can be defined as a function of fast
and slow variables, we propose each order parameter of the amplitude
and phase can be written to a function of a fast and two slow
variables. That is to say, the amplitude and phase of the wave
function are sought for the forms of $A=u_0 (z_0,\xi,\tau)
+\varepsilon[a^{(0)} (z_0,\xi,\tau) +\varepsilon^2 a^{(1)}
(z_0,\xi,\tau)+\cdots]$ and $\varphi=\varepsilon^2[\varphi^{(0)}
(z_0,\xi,\tau) +\varepsilon^2 \varphi^{(1)} (z_0,\xi,\tau)+\cdots]$,
respectively, where the small parameter $\varepsilon$ represents the
relative amplitude of extended states in BEC. Slow variables
$\xi=\varepsilon(z-V_g t)$ and $\tau=\varepsilon^3 t$ characterize
the slow variation of soliton dynamics. Fast variable $z_0=z$
denotes the propagation direction of the lattice wave packets. $V_g$
is a group velocity. By substituting them into Eqs. (\ref{A}) and
(\ref{P}), and then separating them in terms of $\varepsilon$, Eq.
(\ref{A}) can be written as
\begin{equation}
V_g \frac{\partial u_0}{\partial\xi}=0,\label{N3}
\end{equation}
\begin{equation}
V_g \frac{\partial a^{(0)}}{\partial\xi}=\frac{\partial
u_0}{\partial z_0}\frac{\partial \varphi^{(0)}}{\partial z_0}
+\frac{1}{2} u_0\frac{\partial^2\varphi^{(0)}}{\partial z^2_0}
,\label{N4}
\end{equation}
\begin{equation}
\frac{\partial u_0}{\partial z_0} \frac{\partial
\varphi^{(0)}}{\partial\xi} +\frac{\partial u_0}{\partial \xi}
\frac{\partial \varphi^{(0)}}{\partial z_0} +\frac{\partial
a^{(0)}}{\partial z_0} \frac{\partial \varphi^{(0)}}{\partial z_0}
=-\frac{\partial u_0}{\partial\tau},\label{N5}
\end{equation}
\begin{eqnarray}
V_g \frac{\partial a^{(1)}}{\partial\xi} &&= \frac{\partial
a^{(0)}}{\partial\tau} +\frac{\partial a^{(0)}}{\partial \xi}
\frac{\partial \varphi^{(0)}}{\partial z_0} +\frac{1}{2} u_0
\frac{\partial^2\varphi^{(0)}}{\partial \xi^2} \nonumber\\
&&+ \frac{1}{2} u_0 \frac{\partial^2\varphi^{(1)}}{\partial z^2_0}
+a^{(0)}\frac{\partial^2 \varphi^{(0)}}{\partial z_0 \partial\xi}
+\frac{\partial u_0}{\partial z_0} \frac{\partial
\varphi^{(1)}}{\partial z_0} \nonumber\\
&&+\frac{\partial u_0}{\partial \xi} \frac{\partial
\varphi^{(0)}}{\partial \xi} +\frac{\partial a^{(0)}}{\partial z_0}
\frac{\partial \varphi^{(0)}}{\partial \xi}.\label{N6}
\end{eqnarray}
Equation (\ref{P}) becomes
\begin{equation}
-\frac{1}{2}\frac{\partial^2 u_0}{\partial z^2_0} -(\mu-1) u_0
+V_0\sin^2(\frac{\pi z_0}{D}) u_0 +Q' u^3_0=0,\label{N7}
\end{equation}
\begin{eqnarray}
-\frac{1}{2}\frac{\partial^2 a^{(0)}}{\partial z^2_0} &-& (\mu-1)
a^{(0)} +V_0\sin^2(\frac{\pi z_0}{D}) a^{(0)}\nonumber\\
&&+3Q' u^2_0 a^{(0)}=\frac{\partial^2 u_0}{\partial z_0 \partial
\xi},\label{N8}
\end{eqnarray}
\begin{equation}
-\frac{\partial^2 a^{(0)}}{\partial z_0 \partial\xi} +3Q' u_0
(a^{(0)})^2=\frac{1}{2} \frac{\partial^2
u_0}{\partial\xi^2},\label{N9}
\end{equation}
\begin{eqnarray}
-\frac{1}{2}\frac{\partial^2 a^{(1)}}{\partial z^2_0} &-& (\mu-1)
a^{(1)} +V_0\sin^2(\frac{\pi z_0}{D}) a^{(1)} \nonumber\\
&&+3Q' u^2_0 a^{(1)}=\frac{1}{2}\frac{\partial^2 a^{(0)}}{\partial
\xi^2},\label{N10}
\end{eqnarray}
\begin{eqnarray}
\frac{\partial^2 a^{(1)}}{\partial z_0 \partial\xi} &-& 6Q' u_0
a^{(0)} a^{(1)}=\frac{1}{2} u_0
(\frac{\partial\varphi^{(0)}}{\partial z_0})^2\nonumber\\
&&-V_g a^{(0)}\frac{\partial \varphi^{(0)}}{\partial\xi}.\label{N11}
\end{eqnarray}

From Eq. (\ref{N3}), one can see that $u_0$ is independent on $\xi$.
Due to the fast and slow varies possess different physical
connotation, so it is reasonable that these order parameters are
written to arithmetic multiply of function of the fast and slow
variables. We may set $u_0 (z_0, \tau)=u_{01}(z_0) u_{03}(\tau)$.
Similarly, $a^{(i)}$ and $\varphi^{(i)}$ are the forms of $
a^{(i)}(z_0,\xi, \tau)=a_{i1}(z_0)a_{i2}(\xi)a_{i3}(\tau)$ and
$\varphi^{(i)}(z_0,\xi, \tau)
=\varphi_{i1}(z_0)\varphi_{i2}(\xi)\varphi_{i3}(\tau)$,
respectively, where $i=0, 1, 2 \cdots$. Note that the form of Eq.
(\ref{N7}) is the same as that of Eq. (\ref{L2}). In view of the
fact that BEC in the experiments are dilute and weakly interacting:
$n |a_s|^3 \ll 1$, where $n$ is the average density of the
condensate, so Eq. (\ref{N7}) can also be transformed into Mathieu's
equation under the consideration of weak nonlinearity. The solutions
of Mathieu's equation have been discussed in section III. By
comparing Eq. (\ref{N7}) with Eq. (\ref{N8}), one finds $a^{(0)}=0$.
From Eq. (\ref{N4}), we obtain $u_0=(\frac{\partial\varphi^{(0)}}
{\partial z_0})^{-(1/2)}$. So, Eq. (\ref{N5}) becomes

\begin{equation}
-\frac{\frac{\partial \varphi_{03}}{\partial
\tau}}{\varphi^2_{03}}=\frac{\frac{\partial^2\varphi_{01}}{\partial
z^2_0}\varphi_{01}\frac{\partial\varphi_{02}}{\partial\xi}}{\frac{\partial
\varphi_{01}}{\partial z_0}}.   \label{A3}
\end{equation}

The left hand side of Eq. (\ref{A3}) is the differentiation of
$\varphi_{03}$ with respect to $\tau$, while the right hand side
of Eq. (\ref{A3}) is differentiation of $\varphi_{01}$
($\varphi_{02}$) with respect to $z_0$ ($\xi$). Obviously, both
sides of the equation must be equal to a constant $\lambda$, i.e.,
$\varphi_{03}=1/(\lambda \tau)$, and $\varphi_{02}=- (1/2) \lambda
u_{01}\xi [\frac{\partial u_{01}}{\partial z_0} \int\frac{d
z_0}{[u_{01}(z_0)]^2}]^{-1}$. So, $\varphi^{(0)}=-[u_{01}
\xi/(2\tau)] [\frac{\partial u_{01}}{\partial z_0}]^{-1}$.
Correspondingly, we get

\begin{equation}
\frac{\partial u_{03} (\tau)}{\partial \tau} =\frac{u_{03}
(\tau)}{2\tau}. \label{A6}
\end{equation}

Similarly, from Eqs. (\ref{N6}), (\ref{N10}) and (\ref{N11}), we
have

\begin{equation}
a^{(1)} = \frac{u_0 \xi^3}{24 \tau^2 [\frac{\partial^2
u_0}{\partial z^2_0} +4 Q' u^3_0]} \frac{\partial}{\partial
z_0}\{u_0 [1- u_0\frac{\partial^2 u_0}{\partial z^2_0}
(\frac{\partial u_0}{\partial z_0})^{-2}]^2\}.\nonumber\\
~ \label{A9}
\end{equation}

Under the transformations $z_0=z$, $\xi=\varepsilon (z-V_g t)$,
and $\tau=\varepsilon^3 t$, the perturbation parameters can be
written as $ u_0 (z, t)=u_{01}(z) \sqrt{t}$,
$\varepsilon^2\varphi^{(0)}=-[(z-V_g t) u_{0} /(2 t)]
[\frac{\partial u_{0}}{\partial z}]^{-1}$, and $\varepsilon^3
a^{(1)}=\{u_0 (z-V_g t)^3 /[24 t^2 (\frac{\partial^2 u_0}{\partial
z^2} +4 Q' u^3_0)]\} \frac{\partial}{\partial z}\{u_0 [1- u_0
\frac{\partial^2 u_0}{\partial z^2} (\frac{\partial u_0}{\partial
z})^{-2}]^2\}$, where $u_{01} (z)$ is the nonlinear Bloch wave of
BEC in an optical lattice. Finally, the solution of the
dimensionless GP equation (\ref{GP2}) is given by

\begin{eqnarray}
\psi (r,z,t) &=& \sqrt{\frac{1}{\pi}} \exp(-\frac{r^2}{2}) \{ u_{01}
+\frac{u_{01} (z-V_g t)^3}{24 t^2 [\frac{\partial^2 u_{01}}{\partial
z^2} +4 t Q'
u^3_{01}]}\nonumber\\
&& \times\frac{\partial}{\partial z}\{u_{01} [1-
u_{01}\frac{\partial^2 u_{01}}{\partial z^2} (\frac{\partial
u_{01}}{\partial z})^{-2}]^2\}\} \nonumber\\
&&\times\exp[-i \mu t - i\frac{(z-V_g t) u_{01}}{2
t}(\frac{\partial u_{01}}{\partial z})^{-1}].  \label{Wf}
\end{eqnarray}

where the Bloch waves $u_{01}$ and group velocity $V_g$ can be
given in Sec. III. $\mu$ is the chemical potential. Equation
(\ref{Wf}) is just an explicit expression of the wave function for
1D BEC trapped in an optical lattice.

As is discussed in Sec. III, the Bloch wave in the $nth$ energy band
is the linear superposition of the Mathieu's functions $ce_{n-1}$
and $se_{n}$, which is always not satisfy the stable condition of
soliton formation. Therefore, the condensate in the energy band
region can not generate soliton, only the condensates in the band
gaps may be occur the soliton. In the following we discuss soliton
dynamical stabilities of the condensates in the band gaps. The
stability of soliton in nonlinear systems is an important issue,
since only dynamically stable modes are likely to be generated and
observed in experiments.

\subsection{B. Soliton properties in the band gaps}

To link our analytical results to real experiments, we estimate the
values of the dimensionless parameters in Eq. (\ref{Wf}) according
to actual physical quantities. We consider a cigar-shaped $^{87}$Rb
condensate (atomic mass $1.4\times10^{-25}$ $kg$ and the scattering
length $5.3$ $nm$) containing $N=900$ atoms in a trap with
$\omega_z=2\pi\times0.5$ Hz and $\omega_\bot=2\pi\times85$ Hz (The
data are from the experiment \cite{23}). The parameter
$\Omega\approx 0.006\ll 1$ in Eq. (\ref{GP2}). It implies that the
condensate may be regarded as a quasi-1D optical lattice in the
direction of a weak confinement. Hereafter the radial radius is
determined by $r=0.01$. Based on that 1D optical lattice is created
from one pair of counter-propagating laser beams in real
experiments, the lattice depth and the lattice spacing depend on the
peak intensity and the angle of the two identical
counter-propagating laser beams, respectively. For the wavelength
$\lambda_L=783$ $n m$, used in Ref. \cite{23}, this angle between
counter-propagating laser beams would be equal to $2.3^o$. The
periodic potential is $V=E_0 \sin^2 (\pi Z/d)$ where $E_0=0.70$
$E_{rec}$ and $d=\lambda_L/2$ are the lattice depth and periodicity,
respectively. Accordingly, the time and space units correspond to
$0.2$ $ms$ and $3$ $\mu m$, respectively. These units remain valid
for other values of $N$, as one may vary $V_0$ accordingly; in this
case, other quantities, such as $D$, also change. Based on these
proposed, we obtain the dimensionless parameters $D\approx 0.33$,
$Q'\approx 9\times 10^{-3}$ and $V_0\approx7.9$.

As is discussed above, the nonlinear Bloch waves in the $nth$ band
gap can be given by the linear superposition of $ce_n$ and $se_n$
under the case of weak nonlinearity. On the basis of the fact that
the coefficients of periodic Mathieu's functions depend on
eigenvalue $q$ \cite{33} (i.e., $V_0$ in our notation), we
presuppose that the Mahtieu's functions are $ce_1=
\sum\limits^{1000}_{k=0} V_0 \cos(k\eta)$ and $se_1=
\sum\limits^{1000}_{k=0} V_0 \sin(k\eta)$ in the following
calculation.

First, we choose a linear superposition form of $u_{01} (z)=ce_1
+se_1$ as the nonlinear Bloch wave in the first band gap. The
time-evolution of the density distribution of the condensates in
this case is plotted in Fig. 2. We see that the peak of wave packets
decreases exponentially with time and eventually vanishes. Thus the
state is always unstable and be called sub-fundamental soliton in
Ref. \cite{37}. Such a soliton with a very small initial total
number of atoms loses a part of atoms with time going on, so it is
unstable (refer to Ref. \cite{37}).

\begin{figure}[tbp]
\centering
\includegraphics[width=0.5\textwidth]{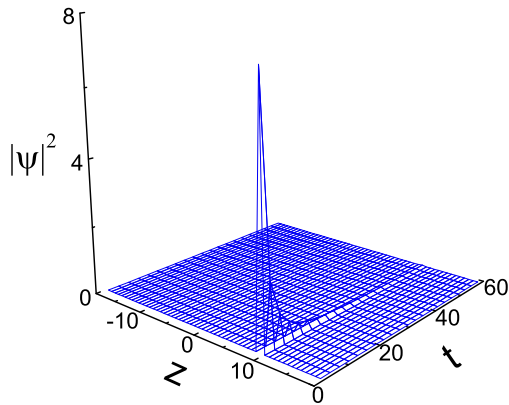}
\hspace{1.0cm} \caption{(Color online). The space-time evolution of
the density of the condensates in the first band gap. The Bloch wave
is chosen as $u_{01} (z)=ce_1 +se_1$. The parameters used are the
lattice depth $V_0=7.9$, interatomic interaction strength
$Q'=9\times 10^{-3}$, the lattice spacing $D=0.33$, the radial
radius $r=0.01$, and the chemical potential $\mu=1.12$. All
parameters are in dimensionless units.}
\end{figure}

\begin{figure}[tbp]
\centering
\includegraphics[width=0.5\textwidth]{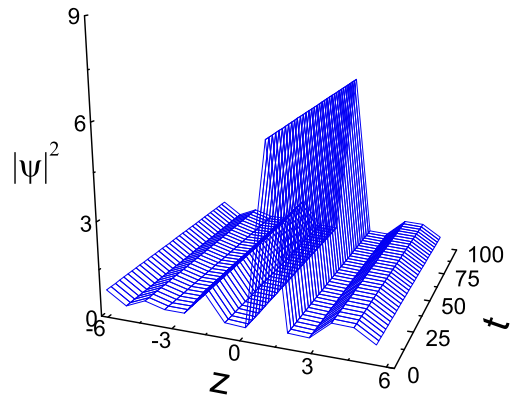}%
\hspace{0.2cm}%
\caption{(Color online). The space-time evolution of the density of
the condensate with $12.6$ dimensionless lengths (about 38 lattice
sites), which corresponds to about $38\mu m$ in real space. The
Bloch wave is chosen as $u_{01} (z)=ce_1 +i se_1$. Other parameters
used are the same as Fig. 2}
\end{figure}

Secondly, we choose $u_{01} (z)=ce_1 +i se_1$, which satisfies the
stable condition of soliton formation, as the nonlinear Bloch wave
in the first band gap. Figure 3 shows the space-time evolution of
the density the condensates in first band gap. A strong peak appears
in the condensate with a dimensionless lengths of 12.6 (about $38$
lattice sites), and maintains its shape and magnitude. It implies
the existence of a bright gap soliton. As the time going on, the
bright gap soliton is pinned in the optical lattice without both
attenuation and change in shape. This behavior indicates that it is
a spatially localized bright gap soliton, which is arisen from the
interplay between the tunneling of periodical potential and
nonlinear interaction of the system. Moreover, the width of the peak
in the $z-t$ plane is found to be a dimensionless length with 2.0,
i.e., $\approx 6$ $\mu m$ in real space. The value is in good
agreement with that of the experiment observed by Eiermann et al.
\cite{23}. The agreement illustrates that our method can describe
the dynamics of BEC trapped in an optical lattice very well. The
type of bright gap solitons are called fundamental solitons in Ref.
\cite{37}. Similar phenomena can also be obtained inside the other
band gap.

From the results discussed above, the solitons residing in the band
gaps are the fundamental or sub-fundamental solitons depending on
their position in the band gap whether the stable condition of
soliton formation can be satisfied or not.

In real experiment, the 1D optical lattice is created from one pair
of counter-propagating laser beams, and the lattice depth depends on
the peak intensity of the two identical counter-propagating laser
beams. That is to say, the lattice depth can be adjusted by varying
the intensity of the counter-propagating laser beams. We here depict
how the lattice depth influences the fundamental soliton in Fig. 4
(with all the rest of the system parameters fixed). From solid line
($V_0=7.9$) and dashed line ($V_0=10$), one sees the amplitude of
the fundamental soliton increasing with the increasing of the
lattice depth. Due to both wave packets containing the same total
number of atoms but the bosons in deeper well are captured more
tightly, the tunneling probability varies smaller \cite{18,niu}. To
balance the same nonlinear effect of the system, the tunneling rate
of bosons in deeper well becomes much larger to achieve the same
dispersion effect, which results in the amplitude of the fundamental
soliton increasing. Therefore, the amplitude of the localized gap
soliton increases with the increasing the intensity of the
counter-propagating lasers beams in the experiments.

\begin{figure}[tbp]
\centering
\includegraphics[width=0.5\textwidth]{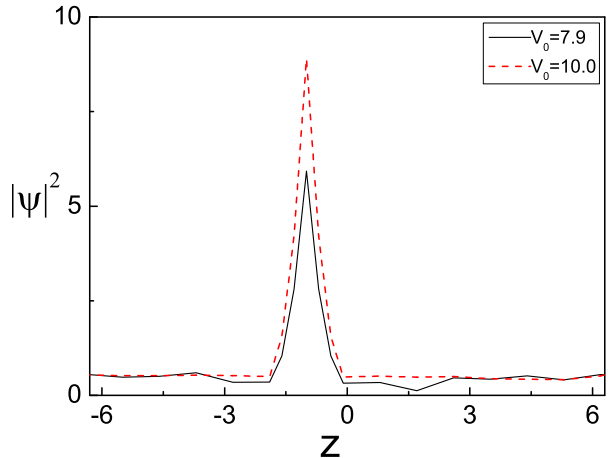}
\hspace{1.0cm} \caption{(Color online). The distribution of the
density of the condensates with different lattice depth at $t=1$,
where the dimensional parameter $V_0=7.9$ corresponds to the
lattice depth in the experiment \cite{23}. Other parameters used
are the same as Fig. 2.}
\end{figure}

Subsequently, we observe the soliton characteristics in a longer
condensate. Our numerical calculations are performed for the
condensate cloud in the ground state extending over 35 dimensionless
lengths (about 106 lattice sites), which corresponds to $105$ $\mu
m$ in real space. The condensate cloud contains about $2.5\times
10^{3}$ atoms under the consideration that the atomic density keeps
unchanged. Fig. 5 shows the space-time evolution of the density of
the condensates in this case. It is shown that there exhibits a
localized gap soliton train consisting of several fundamental
solitons in the condensate. Similarly to the property of a single
fundamental soliton, the solitonlike wave packets in the train are
immobile (i.e., have zero group velocity). In reality, the
condensate is loaded into an optical lattice from a crossed optical
dipole trap \cite{23}, which results in an initial state of a finite
extent. A BEC wave packet centered around a particular quasimomentum
in a given band is created by ramping up of a static lattice with
subsequent linear acceleration to a given velocity \cite{23}. In
order to observe a localized gap soliton train, we may propose
experimental protocols according to the experimental observation of
a single fundamental soliton \cite{23}. First, the atoms are
initially precooled in a magnetic time-orbiting potential trap using
the standard technique of forced evaporation leading to a phase
space density of $\sim 0.03$. Subsequently, the atomic ensemble is
adiabatically transferred into a crossed light beam dipole trap,
where further forced evaporation is achieved by lowering the light
intensity in the trapping light beams. With this approach, one can
generate pure condensates with typically $9\times 10^4$ atoms. By
further lowering the light intensity, one can reliably produce
coherent wave packets of $9000$ atoms. For this atom number no gap
solitons have been observed. Therefore, one removes atoms by Bragg
scattering. This method splits the condensate coherently leaving an
initial wave packet with $2.5\times 10^3$ atoms at rest. Then, the
wave packet centered on a particular position in a given band gap
(which satisfies the stable condition of soliton formation) is
created by switching off one dipole trap beam from a crossed optical
dipole trap, releasing the atomic ensemble into 1D horizontal
waveguide with transverse and longitudinal trapping frequencies
$\omega_\bot =2\pi \times 85$ $Hz$, and $\omega_\| =2\pi \times 0.5$
$Hz$, and then accelerating the periodic potential to the recoil
velocity $v_r=h/m \lambda$. This is done by introducing an
increasing frequency difference between the two laser beams,
creating the optical lattice. The acceleration is adiabatic, which
results in an initial state of an about $105$ $\mu m$ extent. In
view of the fact that the tunneling rate of about 900 atoms
extending the length of $38$ $\mu m$ (in our above simulation) can
balance its nonlinear energy, such a system generates a fundamental
soliton. With both the length of condensate and the total number of
atoms increase, the wave packet exhibits violent dynamics. During
this evolution the wave packet containing $2.5\times 10^{3}$ atoms
is separated from the surrounding atomic cloud into several BEC wave
packets, so the periodic structure of a train of the localized wave
packets emerges. Such a structure represents a train consisting of
several fundamental solitons, which is supported by the combined
action of the repulsive nonlinearity and anomalous diffraction
caused by intersite tunneling in the band gaps \cite{15,17,38}.

\begin{figure}[tbp]
\centering
\includegraphics[width=0.5\textwidth]{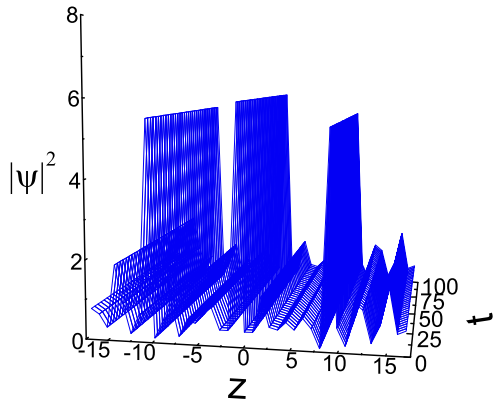}
\hspace{1.0cm} \caption{(Color online). The space-time evolution
of the density of the condensate with 35 dimensionless lengths
(about 106 lattice sites), which corresponds to about $105\mu m$
in real space. Other parameters used are the same as Fig. 2.}
\end{figure}

Finally, we study how the lattice spacing influences the fundamental
soliton or soliton trains as shown in Fig. 6. In practice the
variation of lattice spacing is easy to control by adjusting the
angle between two counter-propagating laser beams. From solid line
($D=0.33$) and dashed line ($D=0.99$), we find that when lattice
spacing occurs a slight difference, the pinning position and the
amplitude of each fundamental soliton have a little change keeping
the distance between adjacent solitons unvaried. It illuminates that
the condensate cloud is separated into three wave packets for both
$D=0.33$ and $D=0.99$. When the lattice spacing $D$ varies from 0.33
to 0.99, the condensate from 91 decreases to 30 wells. Owing to the
center position of each wave packet floating, the pinning position
of the fundamental soliton are set to move. And each BEC wave packet
still forms a localized gap soliton, which comes from the balance
between the nonlinearity and atom dispersion caused by intersite
tunneling. However, when the lattice spacing $D$ varies from $0.33$
to $0.99$, the number of atoms in a well increases from $28$ to
$83$. For the same length condensate, the tunneling between adjacent
well varies easier with the increasing of the atomic number confined
in a well. To balance the same nonlinear energy of the system,
bosons gathering around several wells vary easier, which results in
the amplitude of each localized soliton having a increasing trend.
When lattice spacing varies much larger [see dotted line $D=16.5$ in
Fig. 6], one find that there appears only a fundamental soliton in
the entire condensate with about 30 dimensionless lengths. This
mainly reason is that the condensate in this case only contains
about 2 wells. The tunneling of bosons in adjacent lattice achieves
the dispersion effect to balance the nonlinear energy of the system,
so there exhibits only a fundamental soliton in this case.

\begin{figure}[tbp]
\centering
\includegraphics[width=0.5\textwidth]{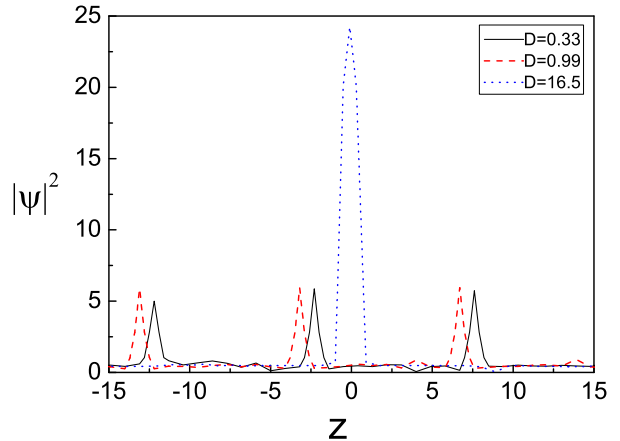}
\hspace{0.01cm} \caption{(Color online). The distribution of the
density of the condensates with different lattice spacing $D$ at
$t=1$, where the dimensional parameter $D=0.33$ corresponds to the
lattice spacing in the experiment \cite{23}. Other parameters used
are the same as Fig. 2.}
\end{figure}

From the results discussed above, we can conclude that the
condensate generating a single fundamental soliton or a localized
gap soliton train consisting of several fundamental solitons can be
controlled by adjusting the length of condensate or (and) the
lattice spacing. Our theoretical results reported here is important
in understanding the fundamental soliton physics of BEC in the
future.

\section{V. CONCLUSION}

In summary, we develop the multiple scale method to study the linear
and nonlinear solitary excitations for 1D BEC confined in an optical
lattice. After averaging over the transverse variable, a
hydrodynamical model of the amplitude and phase is derived. In the
linear case, the Bloch wave in the $nth$ energy band is the linear
superposition of the Mathieu's functions $ce_{n-1}$ and $se_n$, and
the Bloch wave in the $nth$ band gap is the linear superposition of
$ce_n$ and $se_n$. In addition, we find that the stable condition of
soliton formation is that the Bloch wave $u_0$ in $nth$ band gap
satisfies $u_0=\beta ce_n + i\beta se_n$. Under this stable
condition, a novel linear dispersion relation and sound speed are
derived. It is found that the linear dispersion relation and sound
speed depend mainly on the lattice spacing.

For the nonlinear case, we derive a solution of the wave function of
the condensates with weakly inter-atomic interaction, and discuss
its stability for condensate $^{87}Rb$ in band gaps. It shows that
there are two types of gap solitons in the band gaps. One is
fundamental soliton, which is always stable and pins a fixed
position; the other is subfundamental soliton which is always
unstable and decays gradually due to losing a part of its atoms.
Only when the Bloch wave in the band gaps satisfies the stable
condition, the condensates exhibit the fundamental solitons,
otherwise there appears the sub-fundamental solitons. Furthermore,
the pinning position and the amplitude of the fundamental solitons
in the lattice can be controlled by varying the lattice depth and
spacing. We also propose an experimental protocol to observe a
localized gap soliton train consisting of several fundamental
solitons for BEC trapped in an optical lattice in future experiment.

\section{ACKNOWLEDGMENTS}
This work is supported by NSF of China under Grant 90406017,
60525417, 10740420252, 10674070, 10674113, the NKBRSF of China under
Grant 2005CB724508, 2006CB921400, Jiangsu Provincial Postdoctoral
Science Foundation under Grant 0601043B, and Hunan provincial NSF of
China under Grant 06JJ50006.

\end{document}